\documentclass[conference]{IEEEtran}
\IEEEoverridecommandlockouts
\usepackage{cite}
\usepackage{amsmath,amssymb,amsfonts}
\usepackage{graphicx}
\usepackage{textcomp}
\usepackage{xcolor}

\DeclareMathOperator*{\argmin}{argmin}

\usepackage{soul}

\usepackage{algorithm}
\usepackage{algpseudocode}

\usepackage[hidelinks]{hyperref}

\usepackage{amsthm} 
\usepackage{mathtools}

\newtheoremstyle{normalstyle}  
  {}                           
  {}                           
  {\normalfont}                
  {}                           
  {\bfseries}                  
  {.}                          
  { }                          
  {}                           

\theoremstyle{normalstyle}

\newtheorem{Definition}{Definition}

\newtheorem{Theorem}{Theorem}

\newtheorem{Remark}{Remark}

\def\BibTeX{{\rm B\kern-.05em{\sc i\kern-.025em b}\kern-.08em
    T\kern-.1667em\lower.7ex\hbox{E}\kern-.125emX}}
\begin{document}

\IEEEpubid{
\begin{minipage}[t]{\columnwidth}
\begin{flushleft}
\footnotesize\color{red}
This paper has been accepted for presentation at the 23rd European Control Conference (ECC 2025).
\end{flushleft}
\end{minipage}
\hspace{\columnsep}\makebox[\columnwidth]{}
\IEEEpubidadjcol
}

\title{\vspace{0.4 cm} Safety Filter Design for Articulated Frame Steering Vehicles In the Presence of Actuator Dynamics Using High-Order Control Barrier Functions \\
}
\author{Naeim Ebrahimi Toulkani, and Reza Ghabcheloo
\vspace{-0.3 cm} \thanks{Naeim Ebrahimi Toulkani and Reza Ghabcheloo are with the Faculty of Engineering and Natural Sciences, Tampere University,
Korkeakoulunkatu 7, 33720 Tampere, Finland (e-mail: Naeim.EbrahimiToulkani@tuni.fi; Reza.Ghabcheloo@tuni.fi). }
}\vspace{-0.2 cm}

\maketitle
\vspace{-0.2cm}
\begin{abstract}
Articulated Frame Steering (AFS) vehicles are widely used in heavy-duty industries, where they often operate near operators and laborers. Therefore, designing safe controllers for AFS vehicles is essential. In this paper, we develop a Quadratic Program (QP)-based safety filter that ensures feasibility for AFS vehicles with affine actuator dynamics.
To achieve this, we first derive the general equations of motion for AFS vehicles, incorporating affine actuator dynamics.
We then introduce a novel High-Order Control Barrier Function (HOCBF) candidate with equal relative degrees for both system controls.
Finally, we design a Parametric Adaptive HOCBF (PACBF) and an always-feasible, QP-based safety filter. Numerical simulations of AFS vehicle kinematics demonstrate the effectiveness of our approach.
\end{abstract}
\vspace{-0.3 cm}
\vspace{-0.2 cm}
\section{Introduction}
Articulated Frame Steering (AFS) vehicles are widely used in applications where maneuverability is critical, such as off-road environments, construction sites, and other terrains with numerous obstacles and unsafe zones. In these scenarios, the control system must ensure safe navigation, avoiding collisions and other hazards. A key challenge in designing controllers for AFS vehicles is achieving safety while managing the nonlinear dynamics and complex interactions inherent to AFSs.

This challenge becomes more complex when we consider the dynamics of hydraulic actuators, which are commonly used in heavy-duty machinery due to their high power-to-weight ratio and efficiency. In large-scale systems, hydraulic actuators introduce significant dynamics, notably input delays, where the hydraulic response lags behind the commanded control input. Such delays increase the risk of safety violations, as delayed actuation can result in collisions with obstacles.

To address safety challenges, researchers have increasingly employed Control Barrier Functions (CBFs), which are effective for enforcing safety constraints by ensuring system trajectories remain within safe bounds \cite{ames2019control}. For systems with higher relative degrees, CBFs have been extended to High-Order Control Barrier Functions (HOCBFs) \cite{xiao2021high}. 

A common approach to implementing HOCBFs is to use them as safety filters over nominal controllers, applying a minimally modifying Quadratic Program (QP) that adjusts potentially unsafe nominal controls to satisfy HOCBF conditions. However, due to actuation limits in mechanical systems like AFS vehicles, this approach can sometimes lead to infeasibility in the QP. Parameter Adaptive HOCBFs (PACBFs) have been introduced to address these infeasibility issues and enhance reliability \cite{xiao2021adaptive}.
\begin{figure}
\centering
\includegraphics[width=8 cm]{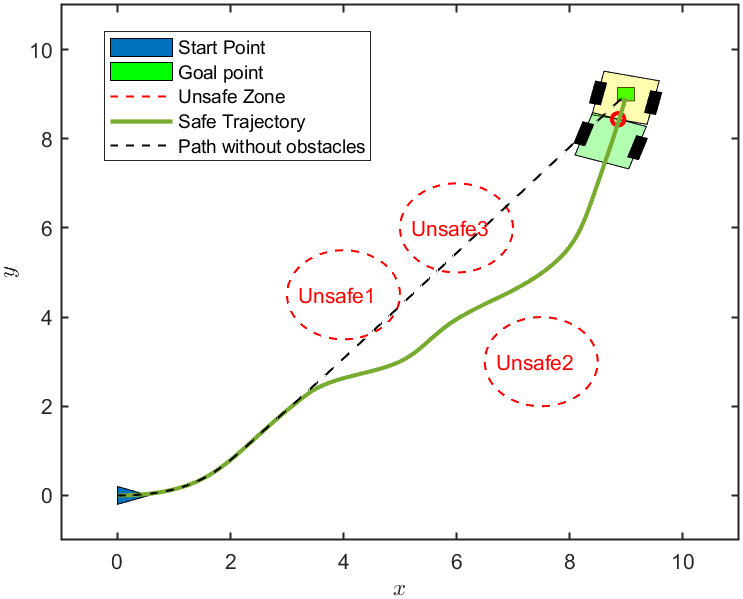}
\vspace{-0.4 cm}
\caption{Safe goal reaching mission for an AFS vehicle considering actuator dynamics using our PACBF-based safety filter.
}
\label{fig:First_fig}
\vspace{-0.5 cm}
\end{figure}

In this paper, we design a safe controller for AFS vehicles using the PACBF framework. Figure~\ref{fig:First_fig} shows the successful application of our developed method on an AFS vehicle in a simulation environment.
The main contributions of this work are as follows:
\begin{itemize}
    \item We derive general equations of motion for AFS vehicles, incorporating actuator dynamics alongside the kinematic equations of AFS vehicles.
    \item We propose a novel HOCBF candidate that maintains equal relative degrees with respect to both control inputs of AFS vehicles.
    \item Based on the proposed HOCBF candidate, we design a QP-based safe controller for AFS vehicles, ensuring feasibility through PACBFs.
    \item We validate our method through the implementation of it on an AFS vehicle in a simulated environment.
\end{itemize}
\vspace{-0.2 cm}
\subsection{Related Works}
AFS vehicles have unique kinematic and dynamic properties due to their central articulation. Early research, such as \cite{corke2001steering}, focused on simplified kinematic models to examine the effects of articulation on path-following and stability.
Delrobaei et al. \cite{delrobaei2011design} designed a Lyapunov function for AFS vehicles, while later studies addressed non-holonomic constraints and the coupling between steering and vehicle dynamics \cite{marshall2008autonomous}.
Despite these efforts, prior works primarily target control without fully addressing the complexities of designing safety controllers for AFS vehicles.

Various studies have examined actuator dynamics and their effects on CBF-based safe controllers. Jankovic \cite{jankovic2018control} introduced time delays to inputs for linear systems, while Seiler et al. \cite{seiler2021control} combined CBFs with integral quadratic constraints (IQCs) to create a robust safety filter. Molnar et al. \cite{molnar2022safety} proposed Environmental Control Barrier Functions (ECBFs) to account for environmental changes during delayed responses, and Zhang et al. \cite{zhang2024eso} developed an extended state observer to anticipate uncertainties and delays within a QP framework.
Additional approaches have tackled actuator dynamics and robustness in safe control. For example, control-dependent barrier functions address low-level dynamics \cite{huang2021safety}, Buch et al. \cite{buch2021robust} designed a robust CBF for uncertain inputs, and Abel et al. \cite{abel2024constrained} presented a state predictor to handle input delays. 
Despite these contributions, a general solution that provides a QP-based safety controller with feasibility guarantees for AFS vehicles remains unaddressed.

\textcolor{black}{
The rest of the paper is organized as follows. Theoretical background and materials are given in Section \ref{section: Preliminaries}. The methodology of our solution is outlined in Section \ref{section: Methodology}. In Section \ref{section:simulation_and_results}, we give simulation results on sample systems and discuss the results. Finally, section \ref{section: Conclusion} concludes the paper.}

\color{black}



\section{Preliminaries and Problem Formulation}
\label{section: Preliminaries}
\color{black}
Consider a nonlinear control-affine system:
\begin{equation}
\label{eq:affine_sys}
\dot{\boldsymbol{x}} = f(\boldsymbol{x}) + g(\boldsymbol{x}) \boldsymbol{u},
\end{equation}
where $\boldsymbol{x} \in \mathbb{R}^{n_x}$ represents the system states, 
$f: \mathbb{R}^{n_x} \to \mathbb{R}^{n_x}$ and $g: \mathbb{R}^{n_x} \to \mathbb{R}^{n_x\times n_u}$ are locally Lipschitz continuous functions, 
and $\boldsymbol{u} \in \mathcal{U} \subset \mathbb{R}^{n_u}$ is the control input. 
The set of admissible inputs $\mathcal{U}$ is defined as
\begin{equation}
    \mathcal{U} = \{\boldsymbol{u} \in \mathbb{R}^{n_u} : \boldsymbol{u}_{min} \leq \boldsymbol{u} \leq \boldsymbol{u}_{max} \}.
\end{equation}

\begin{Definition}
A set of states $X$ is controlled-invariant for a controlled system if, for all initial states in $X$, there exists a control action such that, by its application, the system trajectory remains in $X$ \cite{blanchini1999set}.
\end{Definition}

\begin{Definition}
A continuous function $\alpha: [0,a)\rightarrow [0,\infty)$, with $a>0$, is said to belong to class $\mathcal{K}$ if, it is strictly increasing, and $\alpha(0)=0$ \cite{khalil2002nonlinear}.
\end{Definition}

\begin{Definition}
The Lie derivative of $V$ with respect to $f(\boldsymbol{x})$ is defined as $L_fV(\boldsymbol{x}) = \frac{\partial V}{\partial \boldsymbol{x}} f(\boldsymbol{x})$ \cite{khalil2002nonlinear}.
\end{Definition}

\begin{Definition}
For a (sufficiently many times) differentiable  function $h(\boldsymbol{x}):\mathbb{R}^{n_x} \to \mathbb{R}$, the relative degree of $h$ with respect to $j^{th}$ element of $\boldsymbol{u}$ in \eqref{eq:affine_sys}, is the number of times that we need to differentiate $h(\boldsymbol{x})$ along the dynamics \eqref{eq:affine_sys} until the control element $u_j$ explicitly appears \cite{khalil2002nonlinear}.
\end{Definition}
\begin{Definition}[Control Lyapunov Functions (CLF) \cite{ames2012control}]
A continuously differentiable function $V:\mathbb{R}^{n_x}\to\mathbb{R}$ is an exponentially stabilizing CLF for system \eqref{eq:affine_sys}, if there exist positive constants $c_1,c_2,c_3>0$ such that $\forall\boldsymbol{x}\in\mathbb{R}^{n_x}$:
\begin{subequations}
    \begin{equation}
    \label{eq:CLF_cond_a}
        c_1||x||^2\leq V(\boldsymbol{x}) \leq c_2||x||^2,
    \end{equation}
    \begin{equation}
    \label{eq:CLF_cond_b}
        \inf_{\boldsymbol{u}\in\mathcal{U}} [L_f(V(\boldsymbol{x}))+ L_g(V(\boldsymbol{x}))\boldsymbol{u}+c_3V(\boldsymbol{x})]\leq0.
    \end{equation}
\end{subequations}
\end{Definition}
\subsection{High Order Control Barrier Functions (HOCBF)}
In this paper, we define safety as the positivity of a function, i.e., $h(\boldsymbol{x}) \geq 0$, where $h: \mathbb{R}^{n_x} \to \mathbb{R}$ has a relative degree $m$ with respect to system \eqref{eq:affine_sys}. We set $\psi_0(\boldsymbol{x}) = h(\boldsymbol{x})$ and define a sequence of functions $\psi_i: \mathbb{R}^{n_x} \to \mathbb{R}$ as
\begin{equation}
\label{eq:psi's}
\psi_{i}(\boldsymbol{x}) = \dot{\psi}_{i-1}(\boldsymbol{x}) + \alpha_{i}(\psi_{i-1}(\boldsymbol{x})), \quad i \in \{1, \dots, m\},
\end{equation}
where each $\alpha_i(\cdot)$ is a $(m - i)$th-order differentiable class $\mathcal{K}$ function. Based on \eqref{eq:psi's}, we define the safe sets $\mathcal{C}_1, \dots, \mathcal{C}_m$ as
\begin{equation}
\label{eq:C's}
\mathcal{C}_i = \{ \boldsymbol{x} \in \mathbb{R}^{n_x} : \psi_{i-1}(\boldsymbol{x}) \geq 0 \}, \quad i \in \{1, \dots, m\}.
\end{equation}

\begin{Definition}
\label{definition: HOCBF}
Let $\psi_0(\boldsymbol{x}) = h(\boldsymbol{x})$, and let $\psi_i$ and $\mathcal{C}_i$ be defined by \eqref{eq:psi's} and \eqref{eq:C's}, respectively. The $m^{\text{th}}$-order differentiable function $h: \mathcal{D} \to \mathbb{R}$ is called a High-Order Control Barrier Function (HOCBF) if there exist $(m - i)$th-order differentiable class $\mathcal{K}$ functions $\alpha_i(\cdot)$, $i \in \{1, \dots, m\}$, such that
\begin{equation}
\label{eq:HOCBF}
L_{f}^m h(\boldsymbol{x}) + L_g L_{f}^{m-1} h(\boldsymbol{x}) \boldsymbol{u} + O(h(\boldsymbol{x})) + \alpha_m(\psi_{m-1}(\boldsymbol{x})) \ge 0,
\end{equation}
for all $\boldsymbol{x} \in \mathcal{C}_1 \cap \dots \cap \mathcal{C}_m$. Here, $O(h(\boldsymbol{x}))$ represents the remaining Lie derivatives along $f$ \cite{xiao2021high}.
\end{Definition}

\begin{Theorem} 
\label{theorem:HOCBF}
Let $h: \mathbb{R}^{n_x} \to \mathbb{R}$ be an HOCBF as defined in Definition \ref{definition: HOCBF}. If $\boldsymbol{x} \in \mathcal{C}_1 \cap \dots \cap \mathcal{C}_m$, then any Lipschitz continuous controller $\boldsymbol{u} \in \mathcal{U}$ that satisfies \eqref{eq:HOCBF} renders the set $\mathcal{C}_1 \cap \dots \cap \mathcal{C}_m$ controlled-invariant for system \eqref{eq:affine_sys} \cite{xiao2021high}.
\end{Theorem}

For multi-input systems, one option is to consider the relative degree as the minimum number of times that $h(\boldsymbol{x})$ must be differentiated along the dynamics in \eqref{eq:affine_sys} until any component of the control vector appears in the derivative \cite{xiao2023safe}. 
However, this may limit system performance by restricting the set of control components available to satisfy constraints. In this paper, we focus on the case where we want all control vector components to appear in the HOCBF constraint.


\subsection{Parameter adaptive HOCBFs (PACBFs)}

The notion of HOCBFs is a powerful tool for ensuring safety in dynamical systems. However, the feasibility of optimal control with HOCBF constraints is not always guaranteed. To address this issue, we use PACBFs, which generalize HOCBFs. In PACBFs, we set $\psi_0(\boldsymbol{x}) = h(\boldsymbol{x})$ as the HOCBF and define a series of functions $\psi_i: \mathbb{R}^{n_x} \times \mathbb{R}^{m} \to \mathbb{R}$, $i \in \{1, \dots, m\}$, as
\begin{equation}
\label{eq:psi's2}
\psi_{1}(\boldsymbol{x,p}(t)) = \dot{\psi}_{0}(\boldsymbol{x}) + p_1(t)\alpha_{1}(\psi_{0}(\boldsymbol{x})),
\end{equation}
\begin{equation}
    \begin{multlined}
        \psi_{i}(\boldsymbol{x,p}(t)) = \dot{\psi}_{i-1}(\boldsymbol{x,p}(t)) + \\ 
        p_i(t)\alpha_{i}(\psi_{i-1}(\boldsymbol{x,p}(t))), 
        \quad i = 2, \dots, m,
    \end{multlined}
\end{equation}
where $\alpha_i(\cdot)$, $i \in \{1, \dots, m-1\}$, is a $(m - i)$th-order differentiable class $\mathcal{K}$ function, and $\alpha_m(\cdot)$ is a class $\mathcal{K}$ function. This formulation requires $p_i \geq 0$ ($t$ is omitted for brevity).

Each $p_i$ is defined as an HOCBF with respect to an auxiliary input-output linearizable dynamics:
\begin{equation}
    \label{eq:auxilary1}
    \begin{aligned}
        &\dot{\boldsymbol{\pi}}_i = F_i(\boldsymbol{\pi}_i) + G_i(\boldsymbol{\pi}_i)\nu_i, \quad i \in \{1, \dots, m-1\},\\
        &y_i = p_i,
    \end{aligned}
\end{equation}
where $\boldsymbol{\pi}_i(t) \in \mathbb{R}^{m-i}$ is the vector of auxiliary state variables with $\boldsymbol{\pi}_{m-1}(t)=p_{m-1}(t)\in\mathbb{R}$, $y_i$ denotes the output, $F_i: \mathbb{R}^{m-i} \to \mathbb{R}^{m-i}$ and $G_i: \mathbb{R}^{m-i} \to \mathbb{R}^{m-i}$ are system matrices, and $\nu_i \in \mathbb{R}$ is the control input for the auxiliary dynamics. We then augment \eqref{eq:auxilary1} with dynamics \eqref{eq:affine_sys} to form
\begin{equation}
\label{eq:augmented_dyn0}
\begin{bmatrix} 
    \dot{\boldsymbol{x}} \\ \dot{\boldsymbol{\Pi}}
\end{bmatrix} =
\underbrace{
\begin{bmatrix} f(\boldsymbol{x}) \\ F_0(\boldsymbol{\Pi}) \end{bmatrix}
}_\text{$F(\boldsymbol{x, \Pi})$} +
\underbrace{
\begin{bmatrix}
    g(\boldsymbol{x}) & \boldsymbol{0} \\ \boldsymbol{0} & F_0(\boldsymbol{\Pi}) 
\end{bmatrix}
}_\text{$G(\boldsymbol{x, \Pi})$}
\begin{bmatrix}
    \boldsymbol{u} \\ \boldsymbol{\nu}
\end{bmatrix},
\end{equation}
where $F_0(\boldsymbol{\Pi}) = (F_1(\pi_1), \dots, F_{m-1}(\pi_{m-1}))$, and $G_0(\boldsymbol{\Pi})$ is a matrix composed of $G_i(\pi_i)$, $i \in \{1, m-1\}$.

Since each $p_i$ is an HOCBF candidate with relative degree $(m-i)$ for system \eqref{eq:auxilary1}, we can derive a condition for $\boldsymbol{\nu}$, similar to \eqref{eq:HOCBF}, as follows:
\begin{equation}
\label{eq:U_nu}
\begin{multlined}
\mathcal{U}_\nu(\Pi) = \{ \boldsymbol{\nu} \in \mathbb{R}^{m-1} : L_{F_i}^{m-i}p_i + [L_{G_i}L_{F_i}^{m-i-1}p_i]\nu_i + S(p_i) \\
+ \alpha_{m-i}(\psi_{i,m-i-1}(p_i)) \geq 0, \quad i \in \{1, \dots, m-1\} \},
\end{multlined}
\end{equation}
where $\psi_{i,m-i-1}$ is defined similarly to \eqref{eq:psi's}.
\begin{Remark}
    The exact forms of $F_i$ and $G_i$ do not have a great impact on system performance, as their main role is to ensure the nonnegative property of $p_i$ \cite{xiao2021adaptive}.
\end{Remark}

By assuming that the augmented dynamics \eqref{eq:augmented_dyn0} is linear in $\boldsymbol{\nu}$, we can derive a PACBF condition for the system as
\begin{equation}
\label{eq:PACBF_cond}
\begin{multlined}
\sup_{\boldsymbol{u} \in \mathcal{U}, \  \boldsymbol{\nu} \in \mathcal{U}_{\nu}} 
\left[ L_F^{m}h(\boldsymbol{x}) + [L_G L_F^{m-1}h(\boldsymbol{x})] \boldsymbol{u} \right. \\
\left. + O(h(\boldsymbol{x}), \boldsymbol{p}, \boldsymbol{\nu}) + \alpha_m(\psi_{m-1}(\boldsymbol{x}, \boldsymbol{p})) \right] \geq 0,
\end{multlined}
\end{equation}
where $O(h(\boldsymbol{x}), \boldsymbol{p}, \boldsymbol{\nu})$ is linear in $\boldsymbol{\nu}$ and contains $\alpha_i$, $i \in \{1, \dots, m-1\}$, along with the remaining Lie derivatives along $F$, similar to \eqref{eq:HOCBF}.


\subsection{Kinematic Model of an AFS Vehicle}
\begin{figure}
\centering
\includegraphics[width=8 cm]{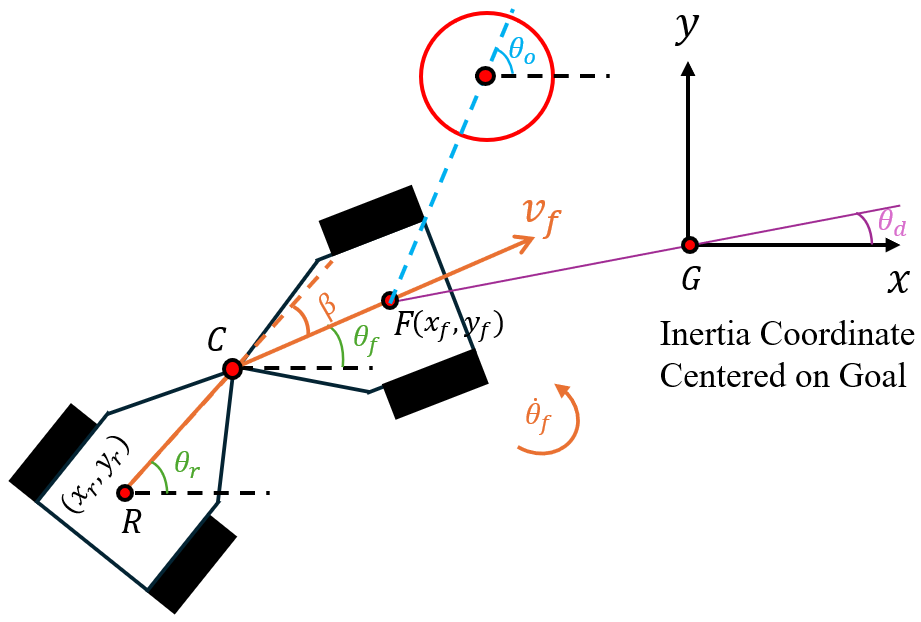}
\vspace{-0.4 cm}
\caption{Schematic of an AFS vehicle and its relative pose to the goal point and an example obstacle}
\label{fig:AFS}
\vspace{-0.5 cm}
\end{figure}

Fig.~\ref{fig:AFS} shows a schematic of an AFS vehicle, where point $C$ is the articulation point and $F(x_f,y_f)$, and $R(x_r,y_r)$ are middle points of the front and rear axels of the machines. Distance from $C$ to $F$ and $C$ to $R$ are denoted as $l_f$ and $l_r$, respectively. 
In addition, $v_f$ is the velocity of the front part, $\beta$ is the articulation angle, and $\theta_f$ and $\theta_r$ are headings of the front and rear units, respectively. 
We adopt the AFS kinematic model proposed in \cite{corke2001steering} to our purpose. Our model of the system is given by:
\vspace{-0.2 cm}
\begin{subequations}
    \label{eq:kinematic_equations}
    \begin{equation}
        \dot{x}_f=v_f \cos{\theta}_f,
    \end{equation}
    \vspace{-0.2 cm}
    \begin{equation}
        \dot{y}_f=v_f \sin{\theta}_f,
    \end{equation}
    \vspace{-0.2 cm}
     \begin{equation}
        \dot{\theta}_f=\frac{v_f \sin{\beta}+l_r\dot{\beta}}{l_f\cos{\beta}+l_r},
    \end{equation}
    \vspace{-0.2 cm}
    \begin{equation}
        \dot{\beta}=u_{\dot{\beta}}.
    \end{equation}
\end{subequations}

\vspace{-0.2cm}
Notice that with state vecotr $\boldsymbol{x}=[x_f,y_f,\theta_f, \beta]^T$ and input $\boldsymbol{u}=[v_{f},u_{\dot{\beta}}]$, system \eqref{eq:kinematic_equations} is a control-affine dynamical system with
\vspace{-0.3 cm}
\begin{equation}
    f_{AFS}(\boldsymbol{x})=
    \begin{bmatrix}
        0 \\ 0 \\ 0 \\ 0
    \end{bmatrix},\ 
    g_{AFS}(\boldsymbol{x})=
    \begin{bmatrix}
        \cos{\theta}_f & 0 \\ \sin{\theta}_f & 0 \\ \frac{\sin{\beta}}{l_f\cos{\beta}+l_r} & \frac{l_r}{l_f\cos{\beta}+l_r} \\ 0 & 1
    \end{bmatrix}.
\end{equation}
\vspace{-0.7 cm}
\section{Methodology}
\label{section: Methodology}
In this section, we design a goal-reaching controller for AFS vehicles to serve as the nominal control input. We then formulate the affine actuator dynamics and augment them to the dynamical system \eqref{eq:kinematic_equations}. Next, we propose an HOCBF candidate for this augmented system and, finally, present a PACBF-based safety filter for AFS vehicles with guaranteed feasibility.
\vspace{-0.25cm}
\subsection{Goal-Reaching Controller Design for AFS Vehicles}
\label{Section:goal_reaching controller} \vspace{-0.1 cm}
We introduce a steering controller to guide the AFS vehicle towards a goal point $G=(x_g, y_g)$. We assume the linear velocity is independently controlled and non-zero, i.e., $v_{f} = v_{ref} \neq 0$, so the vehicle constantly moves forward.

Since the focus of this paper is on safety filter design, we employ a simple performance controller. Without proof and neglecting actuator dynamics, we propose the following kinematic control law:
\vspace{-0.2 cm}
\begin{equation}
\label{eq: Nominal_beta}
    u_{\dot{\beta}_{ref}} = -\frac{v_f}{l_r}\sin{\beta} + \left( \frac{l_f}{l_r} \cos{\beta} + 1 \right) \omega_{ref},
\end{equation}
where \vspace{-0.1cm}
\begin{equation}
    \omega_{ref} = k_\omega(\theta_d - \theta_f),
\end{equation}
\vspace{-0.1 cm}
guides the vehicle towards the line-of-sight angle:
\vspace{-0.1 cm}
\begin{equation}
    \theta_d = \arctan{\frac{y_g - y_f}{x_g - x_f}}.
\end{equation}
\vspace{-0.1 cm}
We denote this control signal as the nominal control inputs $\boldsymbol{u}_N = [v_{ref}, u_{\dot{\beta}_{ref}}]$.

\subsection{AFS Vehicles with Affine Actuator Dynamics}
Assume that an AFS vehicle has affine actuator dynamics given by
\vspace{-0.1 cm}
\begin{equation}
\label{eq:affine_act}
\dot{\boldsymbol{u}} = f_u(\boldsymbol{u}) + g_u(\boldsymbol{u}) \boldsymbol{u}_{cmd},
\end{equation}
\vspace{-0.1 cm}
where $\boldsymbol{u}_{cmd}$ is the input to the system \eqref{eq:affine_act}, and $\boldsymbol{u}$ is its state.

To build a safety controller that accounts for actuator dynamics, we augment system \eqref{eq:affine_act} with \eqref{eq:kinematic_equations}, forming a new dynamical system with state $\boldsymbol{z} = [\boldsymbol{x}, \boldsymbol{u}]^T$, as follows:
\begin{equation}
\label{eq:augmented_dyn}
\dot{\boldsymbol{z}} = 
\underbrace{
\begin{bmatrix} f_{AFS}(\boldsymbol{x}) + g_{AFS}(\boldsymbol{x}) \boldsymbol{u} + f_u(\boldsymbol{u}) \\ \boldsymbol{0} \end{bmatrix}
}_\text{$f_z(\boldsymbol{z})$} +
\underbrace{g_u(\boldsymbol{u})}_\text{$g_z(\boldsymbol{z})$} \boldsymbol{u}_{cmd}.
\end{equation}

In the remainder of this paper, we focus on the dynamical system \eqref{eq:augmented_dyn} and design our safety filter for that.

CBFs have been proposed in the literature for certain kinematic models, and we aim to leverage these existing approaches. For instance, we previously developed a CBF for differential-drive robot kinematics \cite{toulkani2022reactive} with relative degree one in both linear and angular speed inputs. In the next subsection, we extend this approach to AFS vehicles and incorporate actuator dynamics.
\vspace{-0.1 cm}
\subsection{Proposing an HOCBF Candidate for AFS Vehicles}
Safety criteria for AFS vehicles typically depend on maintaining a safe distance from obstacles. A straightforward candidate for this constraint requires the distance from the obstacle’s border to be greater than zero. For a circular obstacle, this can be expressed as
\vspace{-0.2cm}
\begin{equation}
    \label{eq:HOCBF_candidate1}
    h_0(\boldsymbol{z}) = (x_f - x_o)^2 + (y_f - y_o)^2 - R_o^2 \geq 0,
\end{equation}
where $(x_o, y_o)$ and $R_o$ represent the center and radius of the obstacle, respectively. As noted in \cite{toulkani2022reactive} for unicycle vehicles, a key limitation of the HOCBF candidate \eqref{eq:HOCBF_candidate1} is its differing relative degrees with respect to system dynamics, a drawback that also affects AFS vehicles. To address this, we propose the following candidate:
\vspace{-0.1cm}
\begin{equation}
\label{eq:CBF_naeim}
    h_1(\boldsymbol{z}) = (x_f - x_o)^2 + (y_f - y_o)^2 - (R_o + r_s \cos{\eta})^2,
\end{equation}
where $\eta = \min\left(\frac{\pi}{2}, \max\left(-\frac{\pi}{2}, \theta_o - \theta_f\right)\right)$, and $\theta_o$ denotes the angle between the line connecting the vehicle and obstacle centers relative to the x-axis in the inertial frame, as shown in Fig.~\ref{fig:AFS}. Here, $r_s$ is an auxiliary distance parameter that expands the unsafe region by a factor of $\cos{\eta}$.

The HOCBF candidate in \eqref{eq:CBF_naeim} ensures that both control inputs $\boldsymbol{u}_{cmd} = [v_{f_{cmd}}, u_{\dot{\beta}_{cmd}}]$ appear in the HOCBF’s derivatives with respect to the kinematics in \eqref{eq:kinematic_equations} with equal relative degrees. This design allows for more efficient use of both controls to satisfy safety constraints. Equal relative degrees are desirable for multiple-input systems, as an imbalance would introduce the time derivative of one control into the next derivative of the HOCBF, complicating the controller design.

\begin{figure}
\centering
\includegraphics[width=8 cm]{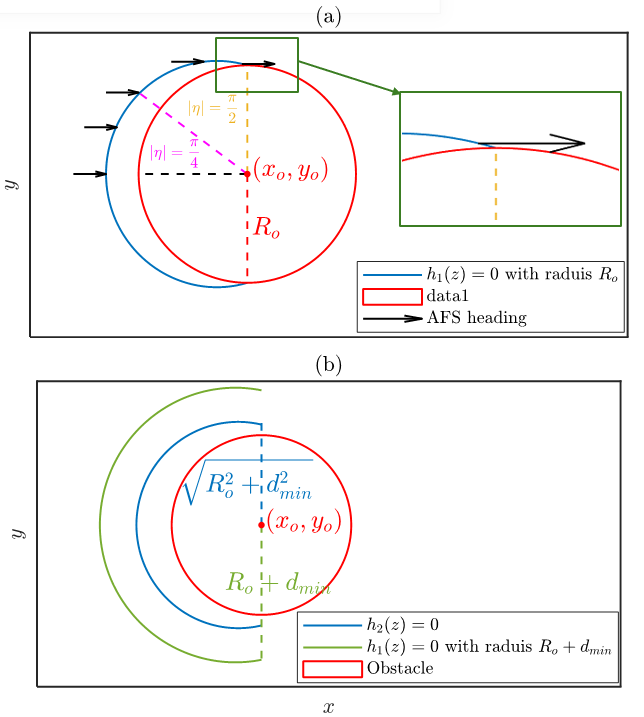}
\vspace{-0.4 cm}
\caption{(a) Effect of formulating safety with $h_1(\boldsymbol{z}) \geq 0$ for different values of $\eta$; (b) Comparison of $h_2(\boldsymbol{z})$ with naively adding a safe distance, $d_{min}$, to the obstacle radius in $h_1(\boldsymbol{z})$.}
\label{fig:rs_plot1}
\vspace{-0.5 cm}
\end{figure}

Although the formulation of the HOCBF candidate in \eqref{eq:CBF_naeim} addresses the relative degree issue, further improvements are possible. First, when $\eta = \pm \frac{\pi}{2}$, the expansion term $r_s \cos{\eta}$ vanishes. In theory, this satisfies safety by treating the system as a point mass. However, in practice, AFS vehicles have considerable size, and we aim to maintain a safe distance from obstacles.
Figure~\ref{fig:rs_plot1}(a) shows the locus of points where $h_1(\boldsymbol{z}) = 0$ for various values of $\eta$ when the vehicle approaches the obstacle with zero heading ($\theta_f = 0$). As $\eta \to \pm \frac{\pi}{2}$, the vehicle may come undesirably close to the obstacle in real-world scenarios. A basic solution is to increase the obstacle size by adding a constant safety margin, replacing the last term of \eqref{eq:CBF_naeim} with $(d_{min} + R_o + r_s \cos{\eta})^2$. This approach improves safety but at the cost of increased conservativeness.
Second, derivatives of a polynomial function like \eqref{eq:CBF_naeim} can approach zero as $h_1(\boldsymbol{z}) \to 0$, requiring larger control inputs $\boldsymbol{u}$ to satisfy \eqref{eq:HOCBF}, which is undesirable.

To address these issues and ensure smoother variations, we propose the following HOCBF candidate:
\vspace{-0.1cm}
\begin{equation}
\label{eq:CBF_naeim2}
    h_2(\boldsymbol{z}) = \ln{\frac{h_1(\boldsymbol{z})}{d_{min}^2}},
    \vspace{-0.1 cm}
\end{equation}
where $d_{min}$ represents the minimum desired distance between the vehicle and the unsafe boundary. Figure~\ref{fig:rs_plot1}(b) shows the locus of points where $h_2(\boldsymbol{z}) = 0$ (i.e., $h_1(\boldsymbol{z}) = d_{min}^2$) for various values of $\eta$ when the vehicle approaches the obstacle with zero heading ($\theta_f = 0$). As illustrated, $h_2(\boldsymbol{z})$ results in a less conservative boundary, allowing the vehicle to come closer to the obstacle when $\eta \to 0$, while maintaining the safe distance $d_{min}$ as $\eta \to \pm \frac{\pi}{2}$.
\vspace{-0.1 cm}
\subsection{Designing a PACBF-Based Safety Filter for AFS Vehicles}
Assume the AFS vehicle receives a nominal control input $\boldsymbol{u}_N$, which may be unsafe in an environment with obstacles. We design an optimal control problem to generate a control trajectory close to $\boldsymbol{u}_N$ while satisfying safety constraints, as follows:
\begin{equation}
    \label{eq:optimal_control}
    \min_{\boldsymbol{u}_{cmd}} \int_0^T R(||\boldsymbol{u}_{cmd}(t) - \boldsymbol{u}_N(t)||),
\end{equation}
where $||\cdot||$ denotes the two-norm, $R(\cdot)$ is a strictly increasing function, and $T > 0$. The optimization in \eqref{eq:optimal_control} must satisfy a safety constraint, $h_2(\boldsymbol{z}) \geq 0$, and input constraints as
\begin{equation}
\label{eq:actuator_lim}
    \mathcal{U}_{cmd} = \{\boldsymbol{u}_{cmd} \in \mathbb{R}^{n_u} : \boldsymbol{u}_{min} \leq \boldsymbol{u}_{cmd} \leq \boldsymbol{u}_{max} \}.
\end{equation}
\vspace{-0.1cm}
To ensure $h_2(\boldsymbol{z}) \geq 0$, we apply the PACBF method. Given that the relative degree of $h_2(\boldsymbol{z})$ with respect to system \eqref{eq:augmented_dyn} is $m = 2$, our PACBF constraints are:
\begin{equation}
    \label{eq:CBF_AFS}
    \begin{aligned}
        &\psi_{1}(\boldsymbol{z,p}(t)) = \dot{h}_2(\boldsymbol{z}) + p_1(t)h_2^2(\boldsymbol{z}),\\
        &\psi_{2}(\boldsymbol{z,p}(t)) = \dot{\psi}_1(\boldsymbol{z}) + p_2(t)\psi_{1}(\boldsymbol{z,p}(t)).
    \end{aligned}
\end{equation}
Note that $\alpha_1(\cdot)$ is a twice-differentiable quadratic function, and $\alpha_2(\cdot)$ is linear. Thus, to maintain system safety, our PACBF condition is $\psi_{2}(\boldsymbol{z,p}(t)) \geq 0$. The design of $h_2(\boldsymbol{z})$ ensures that $v_{cmd}$ and $u_{\dot{\beta}_{cmd}}$ appear in $\psi_{2}(\boldsymbol{z,p})$ in an affine manner. Due to the PACBF construction, $\nu_1$ also appears in $\psi_{2}(\boldsymbol{z,p})$.

In addition, based on the PACBF formulation, we require auxiliary dynamics \eqref{eq:auxilary1} only for $p_1(t)$, which we assume to be linear as follows:
\begin{equation}
    \label{eq:auxilary2}
    \begin{aligned}
        &\dot{p}_1 = \nu_1,\\
        &y_1 = p_1.
    \end{aligned}
\end{equation}
\vspace{-0.1cm}
Since $p_1(t)$ must remain positive, $p_1(t) \geq 0$ serves as an HOCBF for system \eqref{eq:auxilary2} with relative degree one. Following similar steps as in \eqref{eq:CBF_AFS}, HOCBF condition for $\nu_1$ is as
\vspace{-0.1 cm}
\begin{equation}
\label{eq:nu_cond}
    p_1 + \nu_1 \geq 0.
\end{equation}
\vspace{-0.1 cm}
Given that $\nu_1$ has a one-sided constraint, including $\nu_1^2$ in the cost function of our optimal control problem may lead to high values for $p_1$, which contradicts the penalty approach \cite{xiao2021adaptive}. Instead, we minimize $\nu_1$ and add a Control Lyapunov Function (CLF) condition to guide $p_1$ towards a small desired value $p_1^*$. Defining $V_1 = (p_1 - p_1^*)^2$, we derive CLF condition
\begin{equation}
\label{eq:CLF_1}
    2(p_1 - p_1^*)\nu_1 + \epsilon(p_1 - p_1^*)^2 \leq \delta_1,
\end{equation}
where $\epsilon \geq 0$ acts as the coefficient $c_3$ in \eqref{eq:CLF_cond_b}, and $\delta_1$ is a relaxation parameter that makes the CLF condition a soft constraint for \eqref{eq:optimal_control}. To prevent unnecessary deviation from the CLF condition, we include $\delta_1^2$ in the cost function. Similarly, we aim to keep $p_2$ close to a desired value, so we add $(p_2 - p_2^*)^2$ to the cost function. Our revised optimal control problem is now:
\vspace{-0.1 cm}
\begin{equation}
\label{eq:optimal_control2}
\begin{multlined}
    \min_{\boldsymbol{u}_{cmd}, \nu_1, \delta_1, p_2} \int_0^T R(||\boldsymbol{u}_{cmd}(t) - \boldsymbol{u}_N(t)||) + W_1\nu_1 \\
    + P_1\delta_1^2 + Q(p_2 - p_2^*)^2,
\end{multlined}
\end{equation}
subject to \eqref{eq:augmented_dyn}, \eqref{eq:actuator_lim}, \eqref{eq:CBF_AFS}, \eqref{eq:auxilary2}, \eqref{eq:nu_cond}, and \eqref{eq:CLF_1}.

Let $t_f > 0$ be the final time at which \eqref{eq:optimal_control} with constraints \eqref{eq:actuator_lim} and $h_2(\boldsymbol{z}) \geq 0$ remains feasible. The following theorem shows that, under certain assumptions, feasibility of \eqref{eq:optimal_control2} is guaranteed.

\begin{Theorem}
    Suppose $\boldsymbol{z}(0)$ is not on the boundary of the set $\{\boldsymbol{z} \in \mathbb{R}^{n_x + n_u} : \psi_0(\boldsymbol{z}) \geq 0\}$. If the PACBF constraint \eqref{eq:PACBF_cond} is active before $t_f$, then the feasibility of the QP \eqref{eq:optimal_control2} is guaranteed \cite{xiao2021adaptive}.
\label{Theorem:feasibility_guarantee}
\end{Theorem}

Partitioning $[0, T]$ into equal time intervals and assuming constant control within each interval (yielding a piecewise-constant control), we can reformulate \eqref{eq:optimal_control2} as a sequence of QPs. Thus, our problem becomes:
\vspace{-0.1 cm}
\begin{equation}
\label{eq:Safety_Filter}
    \min_{\boldsymbol{u}_{cmd} \in \mathcal{U}, \nu_1, \delta_1, p_2}
    R||\boldsymbol{u}_{cmd} - \boldsymbol{u}_N|| + W_1\nu_1 + P_1\delta_1^2 + Q(p_2 - p_2^*)^2
\end{equation}
\begin{equation*}
    \text{s.t., }
    \eqref{eq:HOCBF}, \eqref{eq:augmented_dyn}, \eqref{eq:auxilary2}, \eqref{eq:nu_cond}, \eqref{eq:CLF_1}.
\end{equation*}

\section{Simulation and Results}
\label{section:simulation_and_results}
We test the designed safety filter on an AFS vehicle in a MATLAB simulation environment. The AFS vehicle is assumed to have affine actuator dynamics defined as:
\begin{subequations}
    \begin{equation}
        \Dot{v}_f = k_1(v_{cmd} - v_f),
    \end{equation}
    \begin{equation}
        \dot{u}_{\dot{\beta}} = k_2(u_{\dot{\beta}_{cmd}} - u_{\dot{\beta}}),
    \end{equation}
\end{subequations}
where $k_1 = k_2 = 4$. The geometric specifications of the simulated AFS vehicle are $l_f = l_r = 1\ \text{m}$, with widths $w_f = w_r = 1\ \text{m}$. The AFS vehicle starts from $X_s = (0,0)$ with zero heading and reaches the goal point $X_g = (9,9)$. For the nominal controller, we use the design in Section \ref{Section:goal_reaching controller} with $v_{ref} = 1\ \text{m/s}$ and $k_w = 1.5$.  

We introduce three circular obstacles centered at $Obs_1 = (4, 4.5)$, $Obs_2 = (7.5, 3)$, and $Obs_3 = (6, 6)$, each with a radius $R_o = 1$. The input constraints $\mathcal{U}_{cmd}$ are defined as:
\begin{equation}
\label{eq:actuator_lim2}
    \begin{bmatrix}
        -1\ \frac{\text{m}}{\text{s}} \\ -23\ \frac{\text{deg}}{\text{s}}
    \end{bmatrix} \leq
    \begin{bmatrix}
        v_{f_{cmd}} \\ u_{\dot{\beta}_{cmd}}
    \end{bmatrix} \leq
    \begin{bmatrix}
        +1\ \frac{\text{m}}{\text{s}} \\ +23\ \frac{\text{deg}}{\text{s}}
    \end{bmatrix}.
\end{equation}

The maximum articulation angle is set to $\beta_{max} = \pm 33\ \text{deg}$. Parameters for the proposed HOCBF are $r_s = \sqrt{2}w_f$ and $d_{min} = \frac{w_f}{2}$, and parameter of CLF \eqref{eq:CLF_1} is set to be $\epsilon = 1$.

The safety filter generates safe controls by solving the QP \eqref{eq:Safety_Filter}, which, given our setup, is specified as:
\begin{equation}
\label{eq:Safety_Filter2}
        \boldsymbol{u}^*(t) = \argmin_{\boldsymbol{u}(t)}
        \frac{1}{2} \boldsymbol{u}^T(t)H\boldsymbol{u}(t) + F^T\boldsymbol{u}(t)
\end{equation}
\begin{subequations}
    \begin{equation*}
        \boldsymbol{u}(t) =
        \begin{bmatrix}
            \boldsymbol{u}_{cmd} \\
            \nu_1 \\ \delta_1 \\ p_2
        \end{bmatrix},\ 
        F = 
        \begin{bmatrix}
            -\boldsymbol{u}_N \\
            W_1 \\ 0 \\ -2Qp_2^*
        \end{bmatrix},
    \end{equation*}
    \begin{equation*}
        H = 
        \begin{bmatrix}
            R_1 & 0 & 0 & 0 & 0 \\
            0 & R_2 &  0 & 0 & 0 \\
            0 & 0 & 0 & 0 & 0 \\
            0 & 0 & 0 & 2P_1 & 0 \\
            0 & 0 & 0 & 0 & 2Q
        \end{bmatrix},
    \end{equation*}
\end{subequations}
subject to \eqref{eq:augmented_dyn}, \eqref{eq:CBF_AFS}, \eqref{eq:auxilary2}, \eqref{eq:nu_cond}, \eqref{eq:CLF_1}, and control bounds \eqref{eq:actuator_lim2}.

We set the coefficients as $R_1 = R_2 = 1$, $W_1 = 1$, $P_1 = 100$, $Q = 100$, $p_1^* = 0.5$, and $p_2^* = 1$. Figure \ref{fig:First_fig} shows the AFS vehicle successfully reaching its goal while avoiding obstacles along its nominal path.

\begin{figure}
    \vspace{0.2 cm}
    \centering
    \includegraphics[width=8 cm]{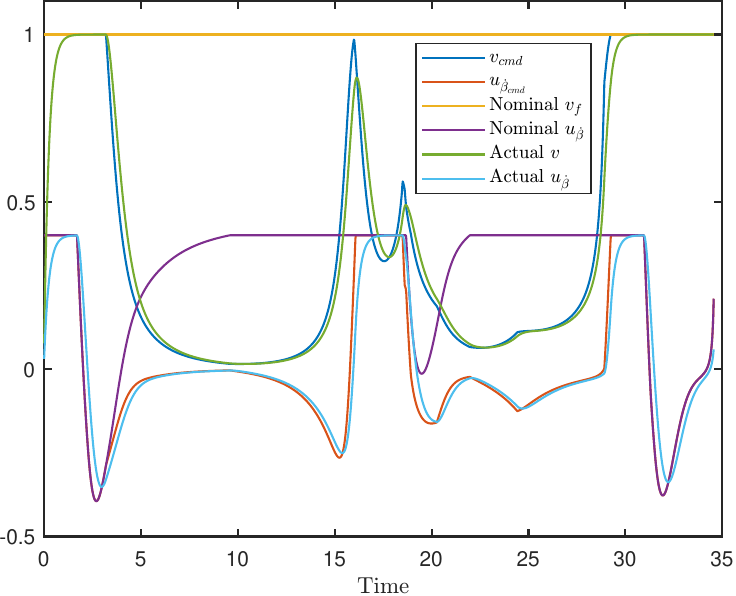}
    \vspace{-0.4 cm}
    \caption{Nominal, command, and actual control input trajectories.}
    \label{fig:control_inputs}
    \vspace{-0.3 cm}
\end{figure}

Figure \ref{fig:control_inputs} shows the nominal, commanded, and actual control inputs. Due to actuator dynamics, actual inputs follow command inputs with a delay. However, the safety filter effectively guides the vehicle to avoid unsafe zones despite these delays.

\begin{figure}
    \centering
    \includegraphics[width=7 cm]{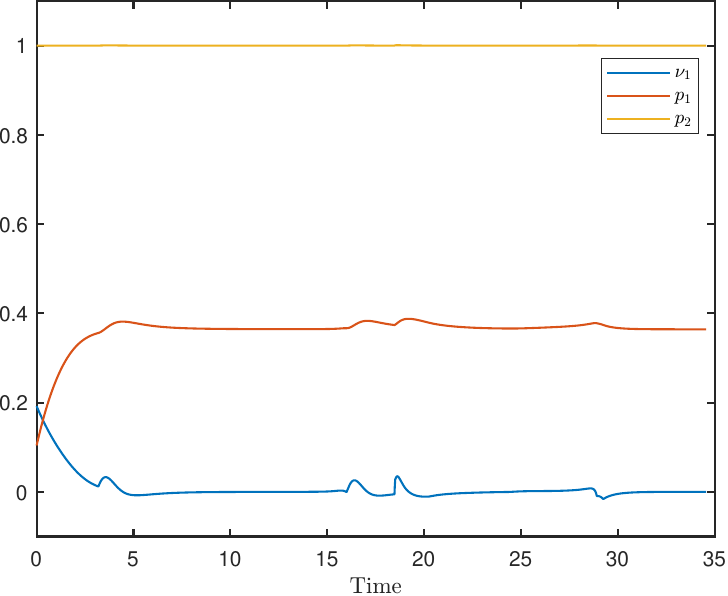}
    \vspace{-0.4 cm}
    \caption{Penalty functions $p_1(t)$ and $p_2(t)$, and dynamics of $\nu_1$.}
    \label{fig:penalties}
     \vspace{-0.5 cm}
\end{figure}

\begin{figure}
\vspace{0.2 cm}
    \centering
    \includegraphics[width=\columnwidth]{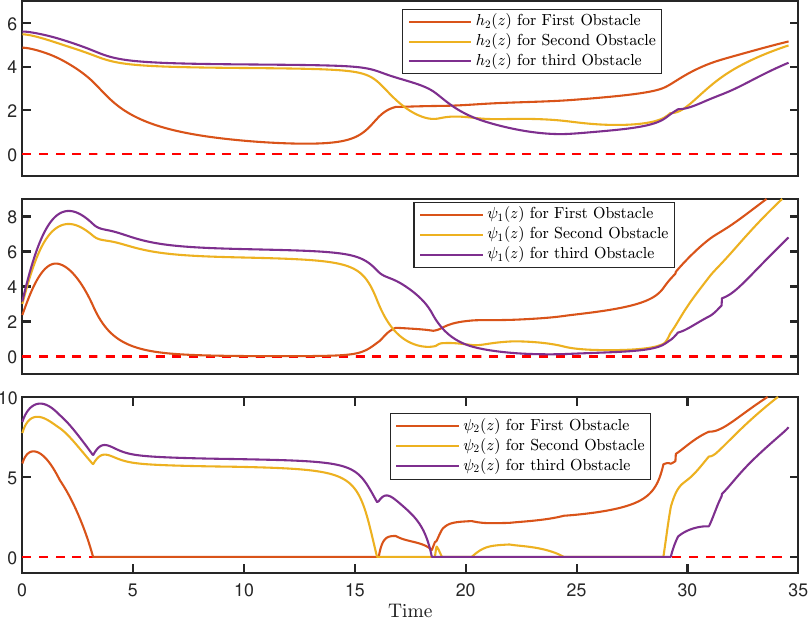}
    \vspace{-0.7 cm}
    \caption{Behavior of HOCBF sequence: $\psi_0(\boldsymbol{z}) = h_2(\boldsymbol{z})$, $\psi_1(\boldsymbol{z})$, and $\psi_2(\boldsymbol{z})$ for different obstacles during the mission.}
    \label{fig:psi_i_positivity}
    \vspace{-0.7 cm}
\end{figure}

The designed safety filter remains feasible by adapting the penalty functions $p_1(t)$ and $p_2(t)$ as needed to maintain QP feasibility. Figure \ref{fig:penalties} illustrates the adjustments in $p_1(t)$, $p_2(t)$, and $\nu_1$ to ensure feasibility of QP \eqref{eq:Safety_Filter2}. The PACBF conditions satisfied by QP \eqref{eq:Safety_Filter2} ensure system safety by keeping the sequence $\psi_i$, $i \in \{0, 1, 2\}$, positive. Figure~\ref{fig:psi_i_positivity} shows the positivity of $\psi_i$, $i \in \{0, 1, 2\}$, throughout the mission.

\section{Conclusion}
\label{section: Conclusion}
This paper presents a safe control design for AFS vehicles with affine actuator dynamics and input limits. We first derive a general navigation formulation for AFS vehicles with actuator dynamics, then apply the notion of HOCBFs by proposing a novel logarithmic HOCBF candidate. Additionally, we enhance the adaptability of the HOCBF through PACBF conditions, ensuring the feasibility of the QP-based safety filter.
The proposed method guarantees safe navigation while accounting for actuator dynamics, and its effectiveness is demonstrated through simulations.

\bibliographystyle{ieeetr}
\bibliography{mybibfile}

\end{document}